\documentstyle[twoside,fleqn,espcrc2,epsf]{article}

\newcommand\figinsert[4]
{\begin{figure}[htb]
\newcommand\figsize{#3}
\epsfysize\figsize
\vskip -.5cm
\centerline{\epsffile{#4}}
\vskip -.5cm
\caption{\leftskip 1pc \rightskip 1pc
\baselineskip=10pt plus 2pt minus 0pt {\sl {#2}}}
\label{fig:#1}
\vskip -.3cm\end{figure}}

\newcommand{\AmS}{{\protect\the\textfont2
  A\kern-.1667em\lower.5ex\hbox{M}\kern-.125emS}}

\def\la{\lambda}

\def\prt{\partial}

\def\vev#1{\langle {#1}\rangle}

\def\frac#1#2{{\textstyle{{#1}\over {#2}}}}

\def\lsim{\mathrel{\rlap{\lower4pt\hbox{\hskip1pt$\sim$}}
    \raise1pt\hbox{$<$}}}
\def\gsim{\mathrel{\rlap{\lower4pt\hbox{\hskip1pt$\sim$}}
    \raise1pt\hbox{$>$}}}
\def\sqr#1#2{{\vcenter{\vbox{\hrule height.#2pt
         \hbox{\vrule width.#2pt height#1pt \kern#1pt
         \vrule width.#2pt}
         \hrule height.#2pt}}}}

\def\CQG{{ Class. Quantum Gravity} }
\def\IJMP{{ Int. J. Mod. Phys.} }

\def\NP{{ Nucl. Phys.} }

\def\PR{{ Phys. Rev.} }

\newcommand{\beq}{\begin{equation}}
\newcommand{\eeq}{\end{equation}}
\newcommand{\bea}{\begin{eqnarray}}
\newcommand{\eea}{\end{eqnarray}}
\newcommand{\rf}[1]{(\ref{#1})}

\title{\Large \bf Dualities, CPT Symmetry and Dimensional
Reduction in String Theory\thanks{Invited talk presented at the Second 
Conference on 
Constrained Dynamcis and Quantum Gravity, Santa Margherita Ligure, 
Italy, September 1996; to appear in Nucl. Phys. B. Suppl.}}

\author{Orfeu Bertolami\address{\it Departamento de F\'\i sica, 
          Instituto Superior T\'ecnico, Av. Rovisco Pais,
          1096 Lisboa Codex, Portugal}}
       
\begin{document}

\begin{abstract}
In this lecture we address the following issues in the context of 
string theories: i) The role played by $S$ and $T$ dualities in obtaining 
topological inflation in N=1 supergravity models, ii) A mechanism to generate 
the baryon asymmetry of the 
Universe based on the  string interactions that violate
CPT symmetry and iii) The quantum cosmology of the dimensionally reduced
multidimensional Einstein-Yang-Mills system.
\end{abstract}

\maketitle

\section{INTRODUCTION}

One of the most challenging aspects of string unification and its D-brane 
relatives is the non-trivial task of reconstructing
our four-dimensional world in a consistent fashion.
The most direct theoretical scenario one can envisage for this purpose is the 
one where the  
important ingredients required for explaning the main features of our 
Universe can be all directly derived from the fundamental theory. This 
implies that
problems such as the initial singularity of classical theories of gravity, 
the cosmological constant problem and the issue of intital conditions of our
Universe among many others can be
all tackled by stringy and/or D-brane physics. The next scenario, 
and  
probably the simplest one from the technical point of view, is to assume 
that for dynamical reasons
the fundamental theory has gone through a considerable dimensional reduction 
process and that solutions of the abovementioned difficulties are to be found 
in the field theory
limit of those theories. Until fairly recently the question whether this
limit was meaningful would prevent one considering 
seriously the latter possibility, however the conjecture that $S$-duality 
\cite{font} is a
symmetry of the fundamental theory gives this hypothesis a better
grounding. Indeed, this symmetry sheds light on the physics of 
the non-perturbative regime and hence on its connection with the 
field theory limit. This implies that 
$S$-duality should, likewise the already well stablished 
$T$-duality \cite{alvarez}, be introduced 
in the supergravity models arising from the fundamental theory. 
It is precisely in the context of these dual
N=1 supergravity models that is shown the moduli fields can be 
stabilized and the dilaton runaway problem solved
\cite{font}. It is also in this context that inflationary models can 
be built, either assuming that the inflaton is
a singlet of the gauge sector \cite{macorra} or that inflation of the
topological type  occurs \cite{bento1}. This particular issue is going to be 
discussed in the first part of this lecture.

In the second part of this lecture, we discuss how a 
sizeable baryon asymmetry of the Universe can be generated via CPT violating 
string interactions \cite{bertolami} arising from the non-trivial solutions 
of the field theory of open strings \cite{kp}.  
Finally, we close with a discussion on the quantum cosmology of 
the multidimensional Einstein-Yang-Mills system after suitable dimensional
reduction \cite{bfm}. We aim with this last 
discussion to illustrate the well known
point that in order to obtain solutions of a given theory a set of initial 
conditions is required and that for this purpose quantum cosmology seems to 
be particularly suitable as it allows for an easy implementation
of the relevant symmetries as well as  for solving the associated 
dynamical contraints.

\section{TOPOLOGICAL INFLATION IN DUAL STRING MODELS}

Superstring unification seems to be particularly fit for constructing 
inflationary cosmological scenarios. The existence
of numerous scalar fields, the moduli, would seem to provide a necessary 
ingredient, however the fact that these fields remain massless at all 
orders in string perturbation theory leads to 
difficulties in  building a 
sucessful cosmological scenario. Among the many moduli, the dilaton ($\phi$),
stands out as it controls the string coupling and variations 
of this field correspond to changes in masses and coupling constants which are
strongly constrained by observation. This question is usually addressed by 
assuming the dilaton develops a potential due to 
non-perturbative effects, such as  gaugino condensation,
and in this way Einstein's gravity can be recovered after $\phi$ settles into 
the minimum of its potential. However, general arguments show that the dilaton
cannot be stabilized in the perturbative regime of string theory leading
to a runaway problem \cite{dine}.
Moreover, even if the dilaton were stabilized by non-perturbative
potentials, these are in general too
steep to be suitable for inflation without fine-tuning the initial conditions 
\cite{binetruy,brustein}.  Of course, once the dilaton is fixed, inflation can 
be achieved via other fields, e.g. chiral fields (gauge singlets) 
for a suitable choice of 
the inflationary sector of the superpotential \cite{ross,bento2}. Furthermore,
there are additional 
difficulties such as 
the Polonyi problem associated with scalar fields 
that couple only gravitationally and which may dominate the energy 
density of the Universe at present 
\cite{coughlan,ellis,bertolami1,carlos,bento4,banks1,banks2,bento5,banks}.

It is known that some of the abovementioned difficulties can be 
avoided imposing the requirement of $S$-duality as it is shown that the 
dilaton potential develops a 
suitable minimum in this case \cite{font}. Already in Ref. \cite{macorra} this 
mechanism was used to stabilise 
the dilaton while inflation was accomplished by chiral fields. 
However, our analysis shows 
that inflation can be achieved via
the dilaton itself provided one considers 
a novel way of implementing the inflationary expansion of the Universe, 
namely topological or defect inflation. 
In this scenario \cite{linde,vilenkin}, one shows that the core of 
a topological defect may undergo exponential inflationary expansion 
provided the scale of symmetry breaking, $\xi$, satisfies the condition

\begin{equation}
\label{aa}
\xi > {\cal O} (M_P)~.
\end{equation}

The ensued inflationary process is eternal since the core of the defect is 
topologically stable and it is the restored symmetry in the core that 
provides the vacuum energy for inflation. We show that the conditions  
for successful inflation are satisfied by domain walls that separate 
degenerate minima 
in $S$-dual superstring potentials \cite{bento1}.

$S$-duality was conjectured \cite{font} in analogy with the 
well-established $T$-duality symmetry of 
string compactification \cite{alvarez}. It is shown that 
the effective 
supergravity action following from string compactification on orbifolds or 
even Calabi-Yau manifolds is constrained by an underlying string 
symmetry, the so-called target space modular invariance \cite{ferrara}. 
The target space modular group $PSL(2,Z)$ acts on the complex scalar field $T$ 
as 

\begin{equation}
\label{ab}
T \to {a T - i b \over i c T + d}\qquad ;a, b, c, d \in Z, ad-bc=1~,
\end{equation}
and $<T>$ is the background modulus associated to the overall scale of 
the internal six-dimensional space on which the string is compactified. 
Specifically, $T=R^2 + i B$, with $R$ being the ``radius'' of the internal 
space and $B$ an internal axion. The target space modular transformation 
contains the well-known duality transformation $R\to 1/R$ as well as discrete 
shifts of the axionic background $B$ \cite{alvarez}. Furthermore, this symmetry
is shown to  
remain unbroken at any order of string perturbation theory. 
The conjectured $S$-duality symmetry would be a further modular 
invariance symmetry in 
string theory, where the modular group now acts on the complex scalar field 
$S=\phi + i \chi $, and $\chi$ is a pseudoscalar (axion) field. 
This symmetry includes a duality invariance under which the dilaton gets 
inverted. $S$-modular invariance strongly constrains the theory 
since it relates the weak and strong coupling regimes as well as the 
``$\chi$-sectors'' of the theory. 

The form of the N=1 supergravity action including gauge and matter fields 
is specified by the functions $G(\Phi, \Phi^*)=K(\Phi, \Phi^*) + 
\ln \vert W(\Phi)\vert^2$ and $f(\Phi)$; $K(\Phi, \Phi^*)$ is the K\"ahler 
potential, $W(\Phi)$ the superpotential and  $f(\Phi)$ the coupling among 
chiral matter fields, generically denoted by $\Phi$, and the kinetic terms
of gauge fields. At string tree-level, the K\"alher potential is given in 
terms of the chiral fields $S,\ T$ by

\beq
\label{ac}
K = -\ln (S + S^*) - 3 \ln (T + T^*)~.
\eeq

The scalar potential can be then written in the following form \cite{font}

\begin{equation}
\label{ad}
V=\vert h^S \vert^2 G_{S S^*}^{-1} + \vert h^T \vert^2 G_{T T^*}^{-1} - 
3 \exp(G)~,
\end{equation}
where $h^i=\exp(\frac{1}{2}G) G^i$, $i=S,T$ and the indices denote 
derivatives with respect to the indicated variable.

We look first at the case where there is only the $T$ modulus field. 
It is known that the purely $T$-dependent superpotential 
has to vanish order by 
order in perturbation theory so that the vacuum expectation value of 
$T$ remains undetermined. 
However, one expects that non-perturbative effects will generate a 
superpotential for $T$. The simplest expression for $W(T)$ compatible with 
modular invariance is \cite{font}

\begin{equation}
\label{ae}
W(T) \sim \eta(T)^{-6}
\end{equation}
and the related scalar potential is given by \cite{font}:

\begin{equation}
\label{af}
V(T) = {1 \over T_R^3\vert \eta(T)\vert^{12}}
\left( {T_R^2 \over 4 \pi^2}\vert \hat G_2(T) \vert^2 -1 \right),
\end{equation}
where $T_R = 2 {\rm Re}\  T$. The function  
$\eta(T)=q^{1/24} \prod_n(1-q^n)$ is the 
Dedekind function, $q = \exp (-2\pi T)$; 
$\hat G_2 = G_2 - 2 \pi/T_R$ is the weight two Eisenstein function 
and $G_2=\frac{1}{3} \pi^2 - 8 \pi^2  \sigma_1(n) \exp(- 2\pi n T)$, where  
$\sigma_1(n)$ is the sum of the divisors of $n$.

If, on the other hand, one considers only the $S$ field, the requirement of 
modular invariance associated with $S$-duality leads to the following 
scalar potential \cite{font}

\begin{equation}
\label{ag}
V(S) = {1 \over S_R\vert \eta(S)\vert^{4}}
\left( {S_R^2 \over 4 \pi^2}\vert \hat G_2(S) \vert^2 - 3 \right).
\end{equation}

This potential (like $V(T)$) diverges in the limits $S\to 0, \infty$ and 
has minima at finite values of $S$, close to the critical value $S=1$. Indeed,
the function $\hat G_2(S)$ has its only zeros at $S=1,\ 
S=\exp(\frac{1}{6} i\pi)$. The potential has other extrema, namely minima, for 
${\rm Re}\ S\sim 0.8, \ 1.3$ and ${\rm  Im}\ S = n,\ n \in Z$;  
these are minima both 
along the ${\rm Re }\ S$ and ${\rm Im}\ S$ directions. The qualitative shape 
of the potential\footnote{Notice that in
order to have a vanishing vacuum energy at the minimum of the potential
a positive cosmological constant has to be added.}
is shown in Figure 1.

\figinsert{fig1}       
{}                     
{2.5truein}{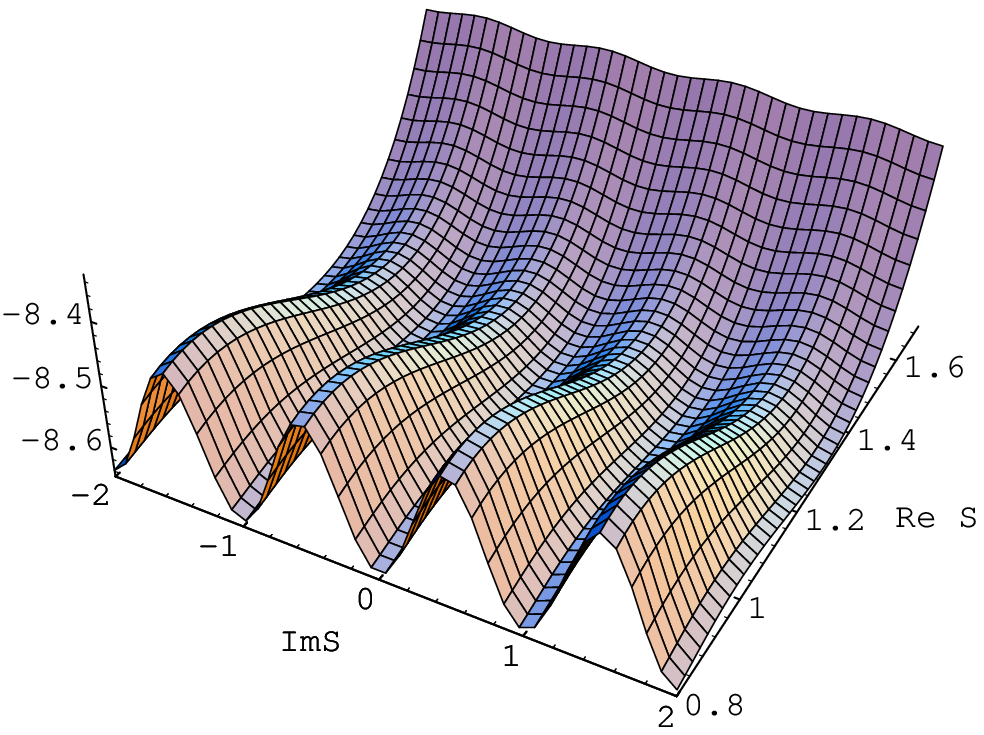}  

Hence, the same way target space modular 
invariance fixes the value of $T_R$ forcing the theory to be 
compactified, $S$-modular invariance fixes the value of $S_R$ thus 
stabilizing the potential and avoiding the dilaton runaway problem. 
Furthermore, it is clear that  the theory has to choose among an infinity 
of degenerate minima whose 
positions differ by modular transformations. Once one of them is chosen, 
target space modular invariance is spontaneously broken. Since duality 
is a discrete symmetry, if there were a phase in the evolution of the Universe 
in which the compactification radius was already
chosen, $S$-duality domain 
walls would be created separating different vacua.

A more realistic model is obtained once all gauge 
singlet fields of the theory: $S, T_i$, $i=1, 2, 3$ are considered and 
$S$ and $T$ dualities are imposed on the 
$N=1$ supergravity Lagrangian \cite{macorra}. A model of this type 
involves also
untwisted chiral fields related to the $T_1$ sector. 
Assuming that  the 
$T$-fields and the untwisted fields of the $T_1$ sector have already settled 
at the minimum of the potential and that inflation takes place due to 
the $S$-field, the relevant potential can be then written as \cite{bento1} 

\begin{equation}
\label{be}
V(S) = {1 \over S_R\vert \eta(S)\vert^{4}}
\left( {S_R^2 \over 4 \pi^2}\vert \hat G_2(S) \vert^2 - a \right),
\end{equation}
where $a$ is a constant. As for the models discussed above, we shall 
further assume that $S_R$ has settled at the minimum of the potential (at
$<S_R>= 2$) and inflation then takes place at the core of domain walls 
that separate different vacua, along the ${\rm Im}\ S$ direction.

In what follows it is shown that the 
conditions for topological inflation to occur at the core of 
the domain walls separating degenerate minima of the  
potential Eq. (8) can be met for a range of the parameter $a$.

Let us first discuss the conditions 
for successful topological inflation. Along a domain wall the field 
$\chi$ ranges from one minimum in one region 
of space to another minimum in another region. Somewhere between, $\chi$ 
must traverse the top of  the potential, $\chi\approx \chi_0$, which 
can be expanded as

\begin{equation}
\label{ah}
V\approx V_0 \left( 1-\alpha^2 {(\chi - \chi_0)^2 \over M^2}\right),
\end{equation}
where $\alpha^2$ is a constant and $M=M_P/\sqrt{8\pi}$, is the natural 
scale of the fields in supergravity
(which has been set to one in the previous discussion).

In flat space, the wall thickness is equal to the curvature of the 
effective potential, that is $\delta^{-1} \sim \alpha(V_0/M_P^2)^{1/2}$. The 
Hubble parameter in the interior of the wall is given by
$H\approx (V_0/3 M^2)^{1/2}$. If 
$\delta \ll H^{-1}$, gravitational effects are negligible. However, if 
$\delta > H^{-1}$, the region of false vacuum near the top of the potential, 
$V\approx V_0$, extends over a region greater than a Hubble volume. Hence, if 
the top of the potential satisfies the conditions for inflation, the interior 
of the wall inflates. Demanding that $\delta > H^{-1}$, one obtains the 
following condition on $\alpha$, 

\begin{equation}
\label{ai}
\alpha^2 < 8 \pi/3~.
\end{equation}

However, the most stringent constraint on $\alpha$ arises for the
requirement that there is at least $N_e$ e-folds 
of inflation (we assume here that $N_e=65$)
\cite{banks}:

\begin{equation}
\label{bb}
\alpha^2 < 3\pi /65~.
\end{equation}

We have computed $\alpha^2$ for the purely $T$-dual and $S$-dual potentials, 
Eqs.
(\ref{af}) and (\ref{ag}), assuming that the real part of the $(T,\ S)$ field 
has already settled at the minimum of the potential. The envisaged scenario 
assumes that inflation takes place at the core of domain walls separating 
different vacua, as the imaginary part of the $(T,\ S)$ field expands 
exponentially once the conditions discussed above are satisfied at the top 
of potential. We find that $\alpha^2=0.30$ and $\alpha^2=0.09$, 
respectively; hence we conclude that the conditions for successful defect 
inflation to occur are fulfilled in the purely S-dual case only \cite{bento1}.
 
In the model with $S$ and $T$ dualities, the value of $\alpha^2$ 
depends on the constant $a$ (and therefore of vacuum contributions 
of the untwisted chiral fields, $\vert P\vert^2$ and $F_0$ as discussed in 
Ref. \cite{bento1}). 
It is found that for $a \ge 2.5 $ (this implies that $F_0 < 0$), 
successful topological inflation
can occur in realistic models as well. We point out that the effect of 
the non-canonical struture of the kinetic terms
of $S$ (and $T$)  dictated by N=1 supergravity , 
$(S_R)^{-2} \partial_\mu S\partial^\mu S^{*}$ 
$((T_R)^{-2} \partial_\mu T\partial^\mu T^{*})$ has been considered 
and it does not affect our results due to the modular invariance.

Hence, we conclude that topological or defect inflation is possible for purely
$S$-dual N=1 supergravity models and in $S$- and $T$-dual   
models where $a \ge 2.5 $ (cf. Eq. (8)), 
thereby solving the cosmological initial condition problems of 
these models.

\section{CPT VIOLATION AND BARYOGENESIS}

Let us now discuss a mechanism for generating the 
baryon asymmetry of the Universe that involves a putative violation 
of the CPT symmetry arising from string interactions \cite{bertolami}. 
The  usual conditions for baryogenesis, namely, violation of baryon number, 
violation of C and CP symmetries, and existence of nonequilibrium processes
are well known
\cite{sakharov}.
These conditions can be met 
in a grand-unified theory (GUT)
through the decay of heavy states at high energy 
\cite{yoshimura,ig,di,we},
through the decay of states in 
supersymmetric or superstring-inspired models 
at somewhat lower energies 
\cite{claudson,ya,be}, 
or via the thermalization of the vacuum energy of 
supersymmetric states \cite{affleck}. 
These conditions can also be satisfied in the electroweak theory
through sphaleron-induced transitions between inequivalent vacua
above the electroweak phase transition 
\cite{kuzmin}.
Spharelon-induced transitions can on the other hand 
dilute baryon asymmetries generated at higher energies 
\cite{kuzmin1}.

Our mechanism is based on the observation that certain string theories may 
spontaneously break CPT symmetry \cite{kp}.
If CPT and baryon number are violated,
a baryon asymmetry could arise 
in thermal equilibrium \cite{dolgov,cohen}. 
A mechanism for baryogenesis along these lines 
has the advantage of being independent 
of C- and CP-violating processes,
which in GUTs are usually rather contrived to account for  
the observed baryon asymmetry and 
are unrelated to the experimentally observed 
CP violation in the standard model.

We assume that the source of baryon-number violation 
is due to processes 
mediated by heavy leptoquark bosons of mass 
$M_X$ in a generic GUT whose details play no essential role
in our mechanism.
Baryon-number violation in the early Universe 
from the leptoquarks is assumed
to be negligible below some temperature $T_D$.

The CPT-violating interactions are shown to arise 
from the trilinear vertex of non-trivial solutions of the field theory 
of open strings and
in the corresponding low-energy four-dimensional effective Lagrangian
via couplings between Lorentz tensors $N$ 
and fermions $\varphi$, $\psi$ \cite{kp}.
Suppressing Lorentz indices for simplicity,
these have the schematic form 
$ L \supset \la {M_S}^{-k} 
N \cdot \overline{\psi} \Gamma (i \partial)^k \varphi + h.c.$,
where $\lambda$ is a dimensionless coupling constant,
$M_S$ is a large mass scale 
(presumably close to the Planck mass),
$\Gamma$ denotes a gamma-matrix structure, 
and $(i\partial)^k$ represents the action of
a four-derivative at order $k \ge 0$.
The CPT violation appears when appropriate components 
of $N$ acquire non-vanishing vacuum expectation values $\vev{N}$.

For simplicity,
only the subset of the CPT-violating terms
leading directly to a 
momentum- and spin-independent energy shift 
of particles relative to antiparticles are considered.
Terms of this form  can generate effects 
in neutral-meson systems that can be observed
\cite{kp,meson}.
These terms are diagonal in the fermion fields 
and involve expectation values  
of only the time components of $N$:
\beq
L = {\lambda \vev{N} \over M_S^k} 
\overline{\psi} (\gamma^0)^{k+1} (i \prt_0)^k \psi + h.c. + ...
\quad .
\label{cptbroken}
\eeq
Since no large CPT violation has been observed,
the expectation value $\vev{N}$ must be suppressed
in the low-energy effective theory.
The suppression factor is presumably
some non-negative power $l$ of 
the ratio of the low-energy scale $m_l$ to $M_S$, that is
$\vev{N} \sim (m_l/M_S)^l M_S$. 
Since each factor of $i\prt_0$
also acts to provide a low-energy suppression,
the condition $k+l = 2$ corresponds to the dominant terms 
\cite{kp}.
In what follows,
we consider the various values of $k$ and $l$ in turn.

For the baryogenesis scenario
it is assumed that each fermion $\psi$ represents 
a standard-model quark of mass $m_q$ and baryon number $1/3$.
The energy splitting between a quark and its antiquark
arising from Eq.\ \rf{cptbroken}
can be viewed as 
an effective chemical potential, $\mu$,
driving the production of baryon number 
in thermal equilibrium. 

We consider a CPT-violating coupling for a single quark field. 
The equilibrium phase-space distributions 
of quarks $q$ and antiquarks $\bar q$ at temperature $T$ are 
$f_q(\vec p)=(1+e^{(E - \mu)/T})^{-1}$ and
$f_{\bar q}(\vec p)=(1+e^{(E + \mu)/T})^{-1}$,
respectively,
where $\vec p$ is the momentum 
and $E = \sqrt{m_q^2 + p^2}$.
If $g$ is the number of internal quark degrees of freedom,
then the difference between the number densities 
of quarks and antiquarks is
\bea
n_q - n_{\bar q} & = &
{g\over(2\pi)^3}
\int d^3 p ~[f_q(\vec p)-f_{\bar q}(\vec p)]
\quad .
\label{barden}
\eea
The contribution to the baryon-number asymmetry 
per comoving volume is given by 
$n_B/s \equiv (n_q - n_{\bar q})/s$, and on its turn
the entropy density $s(T)$ of relativistic particles is given by 
\beq
s(T) ={2\pi^2\over45} g_s(T) T^3
\quad ,
\label{entropy}
\eeq
\beq
g_s (T)=\sum_B g_B\left(T_B\over T \right)^3+
{7 \over 8}\sum_F g_F\left(T_F\over T \right)^3~,
\label{dof}
\eeq
where
the number of degrees of freedom
of relativistic bosons $B$ and fermions $F$
are taken to be $g_B$ and $g_F$,
respectively, such that their component temperatures 
are denoted $T_B$ and $T_F$.
Photon and quark gases have the same temperature $T$.

Considering initially the case $k=0$ with $l=2$, it follows from 
Eq.\ \rf{cptbroken} that the
effective chemical potential is given by
$\mu \sim m_l^2/M_S \simeq 10^{-17} m_l$.
Substitution into Eq.\ \rf{barden} 
and use of the condition $\mu \ll T$,
which is reasonable for any decoupling temperature $T_D$,
gives a contribution to the baryon number per comoving
volume of
\beq
{n_q - n_{\bar q} \over s} \sim
{45g\over{2\pi^4 g_s(T)}}{\mu\over T}I_0(m_q/T)
\quad ,
\label{qcontrib}
\eeq
where 
\beq
I_0(r)=\int_{r}^\infty
dx\,x\sqrt{x^2-r^2}e^x (1+e^x)^{-2}
\quad .
\eeq
This integral satisfies the condition $I_0(r) < I_0(0) = \pi^2/6$.

With two spins and three colours
$g=6$, and the 
result \rf{qcontrib} applies for each flavour.
In GUTs,
$g_s \gsim 10^2$ for $T \gsim 100$ MeV and therefore
the net baryon number per comoving volume
produced with three generations of  
standard-model particles is given by 
$ n_B/s \sim (10^{-2} \mu /T) I_0(m_q/T) 
\sim (10^{-19} m_l /T) I_0(m_q/T)$.
This is however, far too small to reproduce the
observed value $n_B/s \simeq 10^{-10}$.
Notice that  $l\ge 3$ would 
produce even smaller values.
We can therefore exclude 
baryogenesis with standard-model quarks 
via $k=0$ CPT-violating couplings.

We turn now to the cases with $k \ge 1$.
These have CPT-violating couplings 
involving at least one time derivative.
In thermal equilibrium,
it is a good approximation to replace each time derivative
with a factor of the associated quark energy.
This yields energy-dependent contributions 
to the effective chemical potential
given by
\beq
\mu \sim \left({m_l \over M_S}\right)^l {E^k \over M_S^{k-1}}
\quad .
\eeq
Using Eq.\ \rf{barden} it follows that each quark generates a contribution 
to the baryon number per comoving volume of
\beq
{n_q - n_{\bar q} \over s} \sim
{45 g \over 2 \pi^4 g_s(T)} I_k(m_q / T)
\quad ,
\label{basym}
\eeq
where 
\beq
I_k(r) = \int_{r}^\infty dx\, 
{x\sqrt{x^2-r^2} \sinh(\lambda_k x^k)\over\cosh x+\cosh(\lambda_k x^k)}
\quad 
\label{ik}
\eeq
and
\beq 
\la_k = \left({m_l \over M_S}\right)^l 
\left({T \over M_S}\right)^{k-1}
\quad .
\eeq

If $k=1$, 
the dominant contribution arises when $l=1$.
Then,
$\lambda_1 = m_l / M_S \ll 1$
and we have 
\beq 
I_1(r) \approx 
{m_l \over M_S} \int_{r}^{\infty}dx\,
{x^2 \sqrt{x^2 - r^2} \over 1 + \cosh{x}}
\quad .
\eeq
It can be shown that $I_1 < 12m_l/M_S$,
from which implies that the contribution to
$n_B/s$ from the $k=1$ terms is again too small
to reproduce the known baryon asymmetry.

If $k \ge 2$,
the dominant contribution arises when $l=0$.
This gives $\lambda_k = (T/M_S)^{k-1}$. 
Assuming that $T_D << M_S$, 
the integral $I_k$ has integrand peaking near $x \sim 1$
and is exponentially suppressed in the region 
$1 \ll x < M_S/T$.
Moreover, this integral diverges for $x > M_S/T$, but this is 
naturally, an unphysical
artifact of the low-energy approximation.
Since few particles have energy near $M_S$
at temperatures much less than $M_S$,
the integrands can be truncated 
above the region $T \ll E < M_S$ and the integrals become 
\beq
I_k(r) \approx  
\left({T \over M_S}\right)^{k-1}
\int_{r}^{\infty}dx\, {x^{k+1}\sqrt{x^2 - r^2} \over
1 + \cosh{x}}
\quad .
\eeq
This shows that baryogenesis is more suppressed
for $k \ge 2$.

For $k=2$,
$\lambda_2 = T/M_S$. 
A good estimate of the integral $I_2(m_q/T)$ 
can be obtained by setting $m_q/T$ to zero,
since fermion masses either vanish or are much smaller than 
the decoupling temperature $T_D$ and hence 
$ I_2(m_q/T) \approx I_2(0) \simeq 7 \pi^4 T/15 M_S$.
Combining this with Eq.\ \rf{basym}
yields for six quark flavours
a baryon asymmetry per comoving volume given by
\beq
{n_B \over s} \sim
{21 g \over 2 g_s(T)} {T \over M_S}
\simeq {3 \over 5} {T \over M_S}
\quad .
\label{eq:eta}
\eeq
Therefore for an appropriate value of the decoupling temperature $T_D$,
the observed baryon asymmetry of the Universe 
can be obtained provided the interactions 
violating baryon number are still in
thermal equilibrium at this temperature.
In estimating the value of $T_D$,
dilution effects
must naturally be taken into account.
Before discussing these effects we point out that for $k\ge 3$ 
there is an extra suppression by powers of $T/M_S$ which further raises
the decoupling temperature.

Let us now turn to the discussion of processes that can dilute the baryon
asymmetry.
A potentially important source of baryon asymmetry dilution
are the baryon violating sphaleron transitions.
These processes are unsuppressed
at temperatures above the electroweak phase transition
\cite{kuzmin}.

Assuming the GUT conserves the quantity $B-L$, $B$ and $L$
denoting the total baryon- and lepton-number densities,
sphaleron-induced baryon-asymmetry dilution
occurs when $B-L$ vanishes
\cite{kuzmin1}.
This dilution can be estimated by computing the baryon number density 
using standard model fields in thermal equilibrium 
at the temperature $T_S$
when the sphaleron transitions freeze out. 
It is shown that the baryon density is changed due to the sphaleron transitions
to \cite{kuzmin1,bertolami}
\bea
B &=&  \cases{\displaystyle
- {4\over 13\pi^2}\sum_{i=1}^N L_i{m_{l_i}^2\over T^2} 
\quad , 
& 
$B-L =0$
\quad
\cr \displaystyle 
{4 \over 13}(B-L) 
\quad , 
&
$B-L \ne0$
\quad .
\cr}
\label{dilution}
\eea

Considering first the case where $B-L=0$ and
taking the leptoquark decays 
to be dominated by the heaviest lepton of mass $m_L$,
it follows from Eq.\ \rf{dilution} that
the baryon- and lepton-number densities
are diluted through sphaleron effects
by a factor of about $4(N-1)m_L^2/13\pi^2 NT_W^2$.
Combining this result with Eq.\ \rf{eq:eta}
yields a contribution from three generations
to the magnitude of 
the baryon-number asymmetry per comoving volume at present of
\beq
{n_B \over s} \sim
{28 g \over 13 \pi^2 g_s(T_D)}  {m_L^2 T_D \over T_W^2M_S}
\quad .
\label{result}
\eeq
Taking the heaviest lepton to be the tau 
and the freeze-out temperature $T_W$
to be the electroweak phase transition scale,
then baryon asymmetry produced via GUT processes
is diluted by a factor of about $10^{-6}$.
Thus, the observed value of the baryon asymmetry 
can be reproduced if,
in a GUT model where $B-L=0$ initially,
baryogenesis takes place via $k=2$ CPT-violating terms 
at a decoupling temperature $T_D \simeq 10^{-4} M_S$,
followed by sphaleron dilution \cite{bertolami}.
This value of $T_D$ is close to the GUT scale
and leptoquark mass $M_X$,
as is required for consistency.

Notice that in obtaining Eq.\ \rf{result}
we have neglected any possible effects 
from sphalerons occurring at the GUT scale.
The sphaleron transition rate at high temperatures $T$
is $\Gamma \approx \alpha_W^4 T^4$ \cite{ambjorn},
where $\alpha_W$ is the electroweak coupling constant.
From this follows that the rate of baryon-number violation
exceeds the expansion rate of the Universe 
\beq
H\approx \sqrt{g_s} {T^2 \over M_S}
\quad 
\label{H}
\eeq
for temperatures below $\alpha_W^4 M_S \simeq 10^{12}$ GeV
\cite{rs}.
Therefore, sphaleron effects at the GUT scale can 
safely be disregarded.

Of course if Eq.\ \rf{result} is to hold,
then at the GUT scale
the leptoquark interactions that violate baryon number 
must still be in thermal equilibrium.
Assuming baryon number is violated 
via (direct and inverse) leptoquark decays
and scattering,
occurring with gauge-coupling strength $\alpha_X$, 
then
the rates $\Gamma_D$ for decay and 
$\Gamma_S$ for scattering at temperature $T \ge M_X$
are (see, for instance, Ref.\ \cite{we}):
\beq
\Gamma_D = g_s 
{\alpha_X M_X^2 \over \sqrt{T^2 + M_X^2}}~,
\Gamma_S = g_s 
{\alpha_X^2 T^5 \over (T^2 + M_X^2)^2}~. 
\quad
\eeq
It is easy to see that for $T_D \sim M_X$
and a reasonable coupling $\alpha_X$,
both $\Gamma_D/H$ and $\Gamma_S/H $ exceed one
and so the decay and scattering
processes are indeed in thermal equilibrium
at the GUT scale. 

For $B-L\ne0$,
Eq.\ \rf{dilution} shows that there is  
essentially no sphaleron dilution.
However, this is clearly a much less attractive possibility as 
the asymmetry in this case is introduced via initial conditions.
In this situation dilution might occur through other mechanisms
such as, for instance, the dilaton decay in string theories
\cite{yoshimura1,bento}.

In summary, it can be stated that the presence of interactions that 
violate baryon number and of CPT-breaking terms with $k=2$ 
(cf. Eq.\ \rf{cptbroken})
can generate a large baryon asymmetry
with the Universe in thermal equilibrium at the GUT scale.
If the interactions preserve $B-L=0$,
the subsequent sphaleron dilution
reproduces the observed value of the baryon asymmetry \cite{bertolami}.

It is worth remarking that for $k=2$ the decoupling temperature $T_D$
is sufficient low for baryogenesis to be compatible with  
primordial inflationary models of the chaotic type 
and possibly also with new inflationary models.
It is interesting that such models can be built in the context of
superstring models and that these are shown to be consistent with COBE bounds 
on the primordial energy-density fluctuations 
and with the upper bound on the reheating temperature
to prevent the  overproduction of gravitinos 
\cite{ross,bento2}. Our baryogenesis mechanism is in this respect 
also consistent with the topological inflation scenario 
discussed above. This shows that a complete string-based scenario 
can be built such that inflation and baryogenesis can be both 
achieved in a self consistent way.

\section{QUANTUM COSMOLOGY AND DIMENSIONAL REDUCTION}

The issue of dimensional reduction is crucial in string theory as it is in
this process that lies the origin of the effective low-energy field theory. 
In general, theoretical considerations cannot, except in what concerns the 
issue of stability, favour a consistent compactification scheme against any
other. On the other hand, it is the task of phenomenological as well as
cosmological considerations to choose the appropriate ground state of a given
theory. These considerations might still 
turn out to be insufficient to
uniquely select the ground state, but they can be viewed as a guide to rule 
out competing possibilities. A well known example 
in the context of string theory
is the one arising from the requirement that the compactification 
process respect supersymmetry as from that follows that the 
compact six-dimensional manifold is a Calabi-Yau manifold \cite{candelas}. 
One should also require  
that the ground state should, in case 
it is classically stable but semi-classically unstable (we shall discuss 
this issue in the context of our model below), have a life time 
greater than the age of the Universe. 
Vacua degeneracy can of course, be lifted via
identification of symmetries and their breaking. Recent work 
shows that the multiplicity 
of string theories is actually a manifestation of the various sectors of 
a truly more fundamental theory (M-Theory) which are related by duality 
transformations. It would be 
extremely interesting if this theory would allow for uniquely determining its
ground state, although it remains logically possible that only through 
phenomenological and cosmological reasoning this can be achieved. Moreover,
it would be certainly very exiciting to
carry out a quantum cosmology type of analysis in the context of the
fundamental theory.
This would allow for selecting the viable solutions and would be, in the
spirit of the Hartle and Hawking \cite{hh} approach and of the big-fix
mechanism of Coleman \cite{coleman}, the ultimate theory of initial 
conditions.

In the context of string theory the quantum cosmological approach 
has already been considered
for the lowest-order string effective action. This gives origin to  
minisuperspace models where the scale-factor duality,
a special case of $T$-duality for string models embedded in flat homogeneous 
and isotropic manifolds
\cite{veneziano,tseytlin},
can be studied from the quantum mechanical point of view \cite{bento3}
and the possibility of a quantum solution for the graceful exit problem in
pre-big-bang cosmology addressed \cite{gasperini,lukas} 
(as reported by Veneziano in his talk at this conference). Of course, 
one should regard with 
suspicion the procedure of quantizing an effective theory, however it is
possible to argue that this is justifiable as far as the 
truncation of heavy massive modes
is consistent quantum mechanically. Furthermore, given the complexity of the
task in hand, it is certainly quite compelling that already within 
the resulting fairly simple minisuperspace models relevant physical 
issues can be addressed. 
We should add that in what concerns consistency of the procedure 
of quantizing an effective theory, 
coset space compactification and dimensional reduction 
with symmetric fields stands out 
(see Refs. \cite{bmpv,bkm} for a list of the relevant references) 
as this method is shown to be fully consistent 
classically as well as quantum mechanically.
This property of the coset space dimensional reduction 
plays a crucial role in the model we are going to discuss next. 

We consider the minisuperspace model arising from the
coset space compactification of the $D$-dimensional Einstein-Yang-Mills (EYM)
theory \cite{bkm} and report on the result of work done on the quantum
cosmology of this system \cite{bfm}.
This system can be regarded as the bosonic sector 
of string theory once the moduli settle in their ground state, as
discussed in section 2, and some of its features may certainly be
relevant for the understanding of the complete theory. 
We shall see that this simple model
contains many of the features of our previous discussion. 
Our method consists in exploiting the 
isometries of an homogeneous and isotropic spacetime in $D$-dimensions 
to restrict the possible field configurations. This procedure 
gives origin to an effective model with a 
finite number of degrees of freedom where the quantum cosmology 
of the resulting minisuperspace can be examined and the issue 
of compactification addressed. We find that compactifying solutions 
correspond to maxima of the
wave function indicating that these solutions are favoured over the ones
where the extra dimensions are not compactified for an expanding
Universe. We also find that some features of
the wave function of the Universe do depend on the number of extra 
dimensions \cite{bfm}. Before turning to the model, we should mention 
that similar strategy has already 
been applied to obtain the ground-state wave function of the four-dimensional 
EYM system \cite{bm} and when discussing the issue of decoherence 
in the presence of massive vector fields with global symmetries \cite{bmon}.

We consider the $D=d+4$-dimensional EYM action:
\beq
S[\hat g_{\hat\mu\hat\nu}, \hat A_{\hat\mu}] = 
S_{{\rm gr}}[\hat g_{\hat\mu\hat\nu}]
 + 
S_{{\rm gf}}[\hat g_{\hat\mu\hat\nu}, \hat A_{\hat\mu}]
\label{eq:4.1}
\eeq
with
\beq
S_{{\rm gr}}[\hat g_{\hat\mu\hat\nu}]  =  
{1 \over 16 \pi \hat k} \int_{M^D} d \hat{x} 
\sqrt{-\hat g} (\hat R - 2 \hat \Lambda) ~, 
\eeq

\beq
S_{{\rm gf}}[\hat g_{\hat\mu\hat\nu}, \hat A_{\hat\mu}] = 
{1 \over 8\hat{e}^2} 
\int_{M^D} d \hat{x} 
\sqrt{-\hat g} {\rm Tr} \hat F_{\hat\mu\hat\nu} 
\hat F^{\hat\mu\hat\nu} ~,  
\eeq 
where $\hat g = \det 
\left(g_{\hat\mu\hat\nu}\right)$, $\hat R$ is the 
scalar curvature, $\hat k$ and $\hat \Lambda$ are 
the gravitational and cosmological constants in $D-$dimensions, 
$\hat F_{\hat\mu\hat\nu} = 
\partial_{\hat\mu} \hat A_{\hat\nu} - 
\partial_{\hat\nu} \hat A_{\hat\mu} + 
\left[ \hat A_{\hat\mu}, \hat A_{\hat\nu}\right]$ and
$\hat e$ is the gauge coupling constant.

Spacetime admits local coordinates $\hat x^{\hat\mu} =  
(t, x^i, \xi^m)$ -- $\hat\mu = 0,1, \dots, 3+d$; 
$i=1,2,3$; $m=4, \ldots, d+3$, such that:
\begin{eqnarray}
M^D &=& E^{4 + d} = M^4 \times I^d \cr 
    &=& {\bf R} \times G^{{\rm ext}}/H^{{\rm ext}} 
  \times G^{{\rm int}}/H^{{\rm int}}~,
\end{eqnarray}
where $\bf R$ denotes a timelike direction 
and $ G^{{\rm ext}}/H^{{\rm ext}}\left( 
G^{{\rm int}}/H^{{\rm int}}\right)$ the space of external 
(internal) spatial dimensions realized as a coset space of the external 
(internal) isometry group 
$ G^{{\rm ext }}\left(G^{{\rm int}}\right)$.

We consider spatially homogeneous and (partially) 
isotropic field configurations, i.e. symmetric fields (up to gauge 
transformations for the gauge field) under the action of the group 
$G^{{\rm ext }} \times  
G^{{\rm int}}$ and the gauge group $\hat K = SO(N), N \geq 3+d$.

The group of spatial homogeneity and isotropy is: 
\beq
G^{{\rm HI}} = SO(4) \times SO(d+1)~,
\eeq
while the group of spatial isotropy is 
\beq
H^{{\rm I}} = SO(3) \times SO(d)~.
\eeq
We can now consider the alternative realization of $E^{4+d}$ as a coset space: 
\beq
E^{4+d} = {\bf R} \times [SO(4)\times SO(d+1)]/[SO(3) \times SO(d)]~,
\eeq
implying that only $SO(4) \times SO(d+1)$-invariant fields should 
be considered.

The most general $SO(4) \times SO(d+1)$-invariant metric in 
$E^{4+d}$ is the following:
\beq
\hat g = -\tilde N^2(t) dt^2 + \tilde a^2(t) 
\Sigma_{i=1}^3 \omega^i \omega^i 
+ b^2(t) \Sigma_{m=4}^{d+3} \omega^m \omega^m,
\label{eq:4.2}
\eeq
where $\tilde a(t), b(t)$ and the lapse function $\tilde N (t)$ are 
arbitrary non-vanishing functions of time and $\omega^\alpha$ denote 
local moving coframes in $S^3 \times S^d$. 

The $SO(4) \times SO(d+1)-$ symmetric Ansatz for the
gauge field is given by \cite{bkm} (see also Ref. 
\cite{bmpv} for a general discussion on the principles of construction
of this Ansatz):
\begin{eqnarray}
 {\hat A} & = & {1 \over 2}\Sigma_{p,q=1}^{N-3-d}
B^{pq}(t) {\cal T}^{(N)}_{3+d+p\, 3+d+q}dt \nonumber \\
& + & {1 \over 2} \Sigma_{1 \leq i < j \leq 3} 
{\cal T}_{ij}^{(N)} \omega^{ij} \nonumber \\ 
& + & {1 \over 2} \Sigma_{4\leq m < n \leq 3}
  {\cal T}_{mn}^{(N)} \tilde \omega^{m-3\, n-3}\nonumber \\
& + & \Sigma_{i=1}^3 \left[
 {1 \over 4} f_0 (t) \Sigma_{j,k=1}^3  \epsilon_{jik}{\cal T}_{jk}^{(N)}
\right. \nonumber \\
& + & \left. {1 \over 2} \Sigma_{p=1}^{N-3-d} f_p (t) {\cal T}^{(N)}_{i\,
d+3+p} \right] \omega^i \nonumber \\ 
& + & \Sigma_{m=4}^{d+3} \left[ {1 \over 2} \Sigma_{q=1}^{N-3-d} g_q (t) 
 {\cal T}^{(N)}_{m\, d+3+q} \right] \omega^m,
\label{eq:4.3}
\end{eqnarray}
where $f_0(t) f_p (t), p= 1,\ldots, N-3-d; g_q (t), q=1, \ldots, N-3-d; 
B^{pq}(t), 1 \leq p < q \leq N-3-d$ are arbitrary functions and 
${\cal T}_{pq}^{(N)}, 1 \leq p< q \leq N$ are the generators of the gauge 
group $SO(N)$. 
We have used the decomposition 
\begin{eqnarray}
\omega & = & \Sigma_{\alpha = 1}^{d+3} \omega^{\alpha} T_{\alpha}
 + \Sigma_{1 \leq i < j \leq 3} \omega^{ij} {T_{ij}^{(4)} \over 2}
\nonumber \\
& + & \Sigma_{1 \leq m < n \leq d} \tilde \omega^{mn}
{\tilde T_{mn}^{(d+1)} \over 2}  
\end{eqnarray}
for the Cartan one-form in $S^3\times S^d$. Here 
$T_{ij}^{(4)}, \tilde T_{mn}^{(d+1)}$ form a basis of the Lie algebra 
of $G^{{\rm HI}}$, $T_{\alpha} = \frac{T_{\alpha 4}^{(4)}}{2}, 
\alpha = 1,2,3$ and $T_{\alpha} = \frac{T_{\alpha - 3 \, d=1}^{(d+1)}}{2}, 
\alpha = 4, \ldots, d+3$.

Substitution of the  Ans\"atze (\ref{eq:4.2}) and (\ref{eq:4.3}) into 
action  (\ref{eq:4.1}) leads
to the following effective model \cite{bkm}:
\begin{eqnarray}
& S _{\rm eff} &  = 
 16 \pi^2 \int dt N a^3 \Biggl\{ -{3 \over 8\pi k} 
{1 \over a^{2}} \left[{\dot{a} \over N}\right]^2   
\nonumber \\
& + & 
{3 \over 32\pi k} {1 \over a^{2}} + {1 \over 2} \left[
{\dot{\psi} \over N}\right]^2 \nonumber \\
& + & e^{d\beta\psi} {3 \over 4e^2} {1 \over a^{2}} 
\left( {1 \over 2}\left[{\dot{f_0} \over N}\right]^2
+ {1 \over 2}\left[{ {\cal D}_t {\bf f} \over N}\right]^2\right) 
\nonumber \\
& + & e^{-2\beta\psi} {d \over 4e^2} {1 \over b_0^{2}} 
{1 \over 2}\left[{ {\cal D}_t {\bf g} \over N}\right]^2 
\nonumber \\
& - &  W(a,\psi,f_0,{\bf f}, 
{\bf g}) \Biggr\}~, 
\label{eq:4.5}
\end{eqnarray}
where $k = \hat k /v_d b_0^d, e^2 = \hat e^2/v_d b_0^d, 
\beta = \sqrt{16 \pi k / d(d+2)}, v_d$ is the the volume 
of $S^d$ for $b=1$, 
$\psi = \beta^{-1} \ln (b/b_0)$
and ${\cal D}_t$ denotes the covariant derivative with 
respect to the remnant $SO(N-3-d)$ gauge field $\hat B(t)$: 
\beq
{\cal D}_t {\bf f}(t) = {d \over d t} {\bf f(t)} + 
\hat B(t) {\bf f}(t)~, 
\eeq
\beq
{\cal D}_t {\bf g}(t) = {d \over d t} {\bf g(t)} + 
\hat B(t) {\bf g}(t)~, 
\eeq
such that ${\bf f} = \left\{ f_p \right\}$,  
${\bf g} = \left\{ g_q \right\}$  
and $\hat B$ is an
$(N-3-d)\times (N-3-d)$ antisymmetric matrix 
$\hat B = (B_{pq})$. The potential $W$ is on its hand given by: 
\begin{eqnarray}
W & = & e^{-d\beta\psi} \left[ -e^{-2\beta\psi}
{1 \over 16\pi k} {d(d-1) \over 4} {1 \over b_0^{2}} \right.
\nonumber \\
& + & \left. e^{-4\beta\psi} {1 \over b_0^{4}} {d(d-1) \over 8 e^2} 
V_2 ({\bf g})
+  {\Lambda \over 8\pi k}\right] \nonumber \\
& + & e^{-2\beta\psi} {1 \over (ab_0)^{2}} {3d \over 32 e^2} 
({\bf f}\cdot {\bf g})^2 \nonumber \\
& + & e^{d\beta\psi} {3 \over 4 e^2 a^4} 
V_1(f_0, {\bf f})~,
\end{eqnarray}
where 
$\Lambda = v_d b_0^d \hat \Lambda$  and 
\beq
V_1 (f_0, {\bf f})  =   {1 \over 8} 
\left[ \left( f_0^2 + {\bf f}^2 - 1 \right)^2 + 
4 f_0^2 {\bf f}^2 \right],
\eeq
\beq
V_2 ({\bf g})  =  {1 \over 8} \left( {\bf g}^2  - 1\right)^2~,
\eeq
are the contributions associated with the external and internal components 
of the gauge field, respectively.

Introducing the new variables $(\mu, \phi)$: 
$$
a =  \left({k \over 6\pi}\right)^{\frac{1}{2}} e^\mu
$$
$$
\psi =  \left({3 \over 4\pi k }\right)^{\frac{1}{2}} \phi ~,
$$
the canonical conjugate momenta read
$$
\pi_\mu = - \left({2k \over 3 \pi}\right)^{\frac{1}{2}} 
{e^{3\mu} \over N} \dot{\mu}
$$
$$
\pi_\phi = \left({2k \over 3 \pi}\right)^{\frac{1}{2}} 
{e^{3\mu} \over N} \dot{\phi}
$$
$$
\pi_{f_0}  =  {12 \pi^2 \over e^2} e^{d\beta\psi} 
{a \over N} \dot{f}_0
$$
$$
{\bf{\pi_f}}  =  {12 \pi^2 \over e^2} e^{d\beta\psi} 
{a \over N} {\cal D}_t {\bf f}
$$
$$
{\bf \pi_g}  =  {4\pi^2 \over e^2 b_0^2} e^{-2\beta\psi} 
{a^3 \over N} {\cal D}_t {\bf g}~.
$$

The Hamiltonian and $SO(N-3-d)$ gauge constraints which
are obtained by varying the effective action with
respect to $N$ and $\hat B$ respectively, are given by \cite{bfm}:
\begin{eqnarray}
&-&\pi_\mu^2 -  e^{4\mu} + \pi_\phi^2 + 
 e^{2\mu-d\epsilon\phi} {e^2 \over 6\pi^2}
\left[\pi_{f_0}^2 + {\bf \pi_f}^2\right] \nonumber \\
&+& e^{2\epsilon\phi} {3 e^2 b_0^2 \over d \pi k} 
{\bf \pi_g}^2 +  e^{6\mu} \left({4 k \over 3}\right)^2W = 0~,
\end{eqnarray}
\vskip 0.1cm
\beq
\pi_{f_p} f_q + \pi_{g_p} g_q 
- \pi_{f_q} f_p - \pi_{g_q} g_p = 0~,
\eeq
where $\epsilon = \sqrt{12/d (d+2)}$.

The canonical quantization procedure follows by replacing the canonical
conjugate momenta into operators,
$ \pi_\mu \mapsto -i\frac{\partial}{\partial \mu}$,
$\pi_\phi \mapsto -i\frac{\partial}{\partial \phi}$, etc, and
the resulting Wheeler-DeWitt equation for the wave function
$\Psi=\Psi(\mu, \phi, f_0, {\bf f}, {\bf g})$ is, 
after setting $p=0$ when parametrizating the operator order ambiguity 
$\pi_\mu^2 \mapsto - \mu^{-p} \frac{\partial}{\partial \mu} 
\left(\mu^p  \frac{\partial}{\partial \mu} \right)$, given by \cite{bfm}: 
\bea
&& \Bigl\{{\partial^2 \over \partial \mu^2} 
-  e^{4\mu} - {\partial^2 \over \partial \phi^2} \nonumber \\
& - & e^{2\mu-d\epsilon\phi} {e^2 \over 6\pi^2}
\left[{\partial^2 \over \partial f_0^2}
+ {\partial^2 \over \partial {\bf f}^2} \right] - e^{2\epsilon\phi} 
{3 e^2 b_0^2 \over d \pi k} 
{\partial^2 \over \partial {\bf g}^2} \nonumber \\
& + & e^{6\mu} \left({4 k \over 3}\right)^2 W\Bigr\} \Psi = 0~.
\eea
 
\vskip 0.2cm

To study the compactification process we set the gauge field to its 
static vacuum configuration:
$$
~f_0 = f_0^v, ~{\bf f} = {\bf f}^v, ~{\bf g} = {\bf g}^v = 
{\bf 0}~.
$$
Furthermore, we also assume that {\bf f} and {\bf g} 
are orthogonal. 

Setting 
\beq
v_1 \equiv V_1 (f_0^v, {\bf f}^v)
\qquad
v_2 \equiv V_2 ({\bf g}^v)={1 \over 8}~~,
\eeq
our problem simplifies to
\beq
\left[{\partial^2 \over \partial \mu^2} 
-  {\partial^2 \over \partial \phi^2} + U(\mu, \phi) \right] \Psi(\mu,\phi) 
= 0,
\eeq
where
\beq
U(\mu, \phi) = e^{6\mu}  \left({4 k \over 3}\right)^2 
\Omega (\mu, \phi) - e^{4 \mu}
\eeq
and
\begin{eqnarray}
\Omega (\mu, \phi) & = & e^{-d\epsilon\phi} \left[ -e^{-2\epsilon\phi}
{1 \over 16\pi k} {d(d-1) \over 4} {1 \over b_0^{2}} \nonumber 
\right. \nonumber \\ 
& + & \left. e^{-4\epsilon\phi} {1 \over b_0^{4}} {d(d-1) \over 8 e^2} 
v_2 + {\Lambda \over 8\pi k}  
\right] \nonumber \\
&+& e^{d\epsilon\phi -4 \mu} \left({6\pi \over k}\right)^2 
{3 \over 4 e^2 } v_1 ~.
\end{eqnarray}

Stability of compactifation requires, as discussed in Ref. \cite{bkm}, that
the $D$-dimensional cosmological constant satisfies the condition:
\beq
{c_1 \over 16\pi k}<\Lambda< {c_2 \over 16\pi k}~,
\eeq
where $c_1=d(d-1)e^2/16v_2$ and $c_2=[(d+2)^2(d-1)/(d+4)]e^2/16v_2$.

On the other hand, in order to match the observational bound, 
$|\Lambda^{(4)}|<10^{-120}\frac{1}{16\pi k}$, it is required that  
\beq
\Lambda={d(d-1) \over 16b_0^2}
\eeq
where $b_0^2 = 16\pi k v_2/e^2$. This condition ensures, 
at vanishing temperature \cite{bkm}, the classical
as well as the semiclassical stability of the vacuum.

The potential in the minisuperspace simplifies then to:
\begin{eqnarray}
U(\mu,\phi) &=& e^{6 \mu-d \epsilon \phi} {2k \Lambda \over 9 \pi}
\left( e^{-2\epsilon\phi}-1 \right) ^2-e^{4\mu} \nonumber \\
&+& e^{2\mu+d \epsilon
\phi} {3\pi \over k} {v_1 \over v_2}b_0^2 ~.
\end{eqnarray}

Next we consider the Hartle-Hawking path integral representation 
for the ground-state wave function of the Universe \cite{hh}
\beq
\Psi[\mu,\phi]=\int_{C}{D\mu D\phi \exp(-S_{\rm E})}
\eeq
which allows to evaluate the solution of (49), $\Psi(\mu,\phi)$, 
close to $\mu=-\infty$
($S_{\rm E}=-iS_{\rm eff}$ and C is a compact
manifold with no boundary) and to stablish  the regions where the wave 
function behaves as an exponential (quantum regime) or as an oscillation 
(classical regime). The details of this analysis can be found in Ref.
\cite{bfm}. Our results indicate that a generic feature of the 
wave function is that solutions 
corresponding stable compactifying solutions are maxima, that is the
most probable configurations, for an 
expanding Universe as shown, for instance in Figure 2. Moreover,
some properties of the wave function were found to depend on the number, 
$d$, of internal space dimensions \cite{bfm}.

\figinsert{fig2}
{}                
{2.4truein}{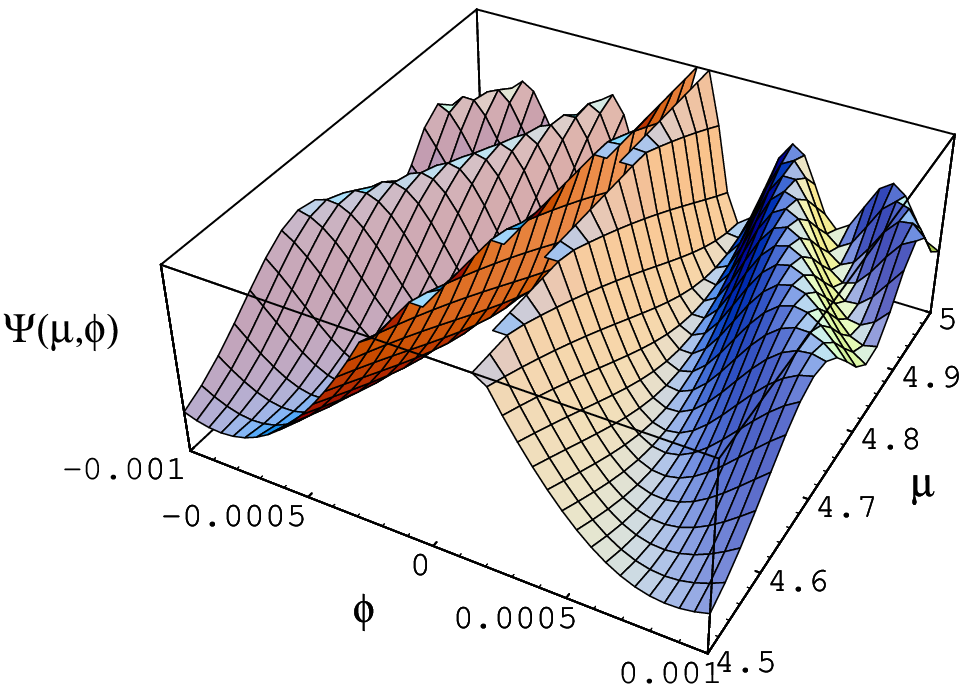}

We stress that a distinctive feature of our scheme is the non-vanishing 
contribution of the external components of the gauge field
to the potential $V_1$ in $W$. 
It is precisely this feature which makes possible to obtain an absolute
classically stable compactification 
after inflation \cite{bkm}
and that is responsible for some of the dependence of the wave function in the 
number, $d$, of internal dimensions \cite{bfm}. 
It is also in this respect that our work contracts with previous one in 
the literature such as for instance the  
6-dimensional Einstein-Maxwell model
with an ``internal'' magnetic monopole discussed in Ref. \cite{jh}
and the model where gravity is coupled with an ``internal'' 
$(D-4)$th rank antisymmetric tensor of Ref. \cite{cw}. 
Our results are although  more general,
consistent with the ones of the those 
references in the particular situation where the external components of the 
gauge field vanish. 

We summarize our conclusions as follows. The issue 
of stability of solutions and the presence of correlations
in the quantum case can only be properly addressed via  Ans\"atze
for the fields that account for the dynamical degrees of freedom
consitently with internal and spacetime symmetries.  
Moreover, our results demonstrate that once 
the four-dimensional cosmological constant is constrained 
to is observational upper bound and the corresponding 
$D$-dimensional one is chosen in order to allow the stability of the internal 
dimensions, 
then the wave function clearly exhibits a correlation 
between compactification of internal dimensions and expansion of 
the external ones. This illustrates
how a phenomenological requirement help in choosing the ground state of the 
theory in hand and the way quantum cosmology may be regarded as yet 
another procedure to select the vacuum of unification theories describing 
our four-dimensional world.

\vskip 0.3cm

It is a pleasure to thank Pedro D. Fonseca for a careful reading of the 
original manuscript and for many relevant suggestions. I would like also
to express my gratitude to the organizing committee of the Second Conference
on Constrained Dynamics and Quantum Gravity and, in particular, to Vittorio De 
Alfaro and Marco Cavagli\`a for the friendship over the years and for 
the warm hospitality in Santa Margherita.


\begin{thebibliography}{99}

\bibitem{font} A. Font, L.E. Ib\'a\~nez, D. L\"ust and F. Quevedo, Phys. Lett.
{\bf B245} (1990) 401; {\bf B249} (1990) 35.

\bibitem{alvarez} E. Alvarez and M. Osorio, Phys. Rev. {\bf D40} (1989) 1150.

\bibitem{macorra} A. De La Macorra and S. Lola, Phys. Lett. {\bf B373} (1996) 
299. 

\bibitem{bento1} M.C. Bento and O. Bertolami, Phys. Lett. {\bf B384} (1996) 98.

\bibitem{bertolami} O. Bertolami, D. Colladay, V.A. Kosteleck\'y
and R. Potting, ``CPT Violation and Baryogenesis'' (hep-ph 9612437); to 
appear in Phys. Lett. B.

\bibitem{kp} 
V.A. Kosteleck\'y and R. Potting, 
Nucl. Phys.  {\bf B359} (1991) 545;
Phys. Rev.  {\bf D51} (1995) 3923; 
Phys. Lett.  {\bf B381} (1996) 389. 

\bibitem{bfm} O. Bertolami, P.D. Fonseca and P.V. Moniz, ``Quantum 
Cosmological Multidimensional Einstein-Yang-Mills Model in a $R \times 
S^3 \times S^{d}$ Topology'' (gr-qc 9607015).

\bibitem{dine} M. Dine and N. Seiberg, Phys. Lett. {\bf B162} (1986) 299.

\bibitem{binetruy} P. Bin\'etruy and M.K. Gaillard, Phys. Rev. {\bf D34} 
(1986) 3069.

\bibitem{brustein} R. Brustein and P.J. Steinhardt, Phys. Lett. 
{\bf B302} (1993) 196.

\bibitem{ross} G.G. Ross and S. Sarkar, Nucl. Phys. {\bf B461} (1996) 597.

\bibitem{bento2} M.C. Bento and O. Bertolami, Phys. Lett. {\bf B365} (1996) 59.

\bibitem{coughlan} G.D. Coughlan, W. Fischler, E.W. Kolb, S. Raby and
  G.G. Ross, Phys. Lett. {\bf B131} (1983) 59.

\bibitem{ellis} J. Ellis, D.V. Nanopoulos and M. Quir\'os, 
Phys. Lett. {\bf B174} (1986) 176.

\bibitem{bertolami1} O. Bertolami, Phys. Lett. {\bf B209} (1988) 277.

\bibitem{carlos} B. de Carlos, J.A. Casas, F. Quevedo and E. Roulet, 
Phys. Lett. B318 (1993) 447.

\bibitem{bento4} M.C. Bento and O. Bertolami, Phys. Lett. {\bf B336} (1994) 6.

\bibitem{banks1} T. Banks, D.B. Kaplan and A.E. Nelson, 
 Phys. Rev. {\bf D49} (1994) 779.

\bibitem{banks2} T. Banks, M. Berkooz and P.J. Steinhardt, 
Phys. Rev. {\bf D52} (1995) 705.

\bibitem{bento5} M.C. Bento and O. Bertolami, Gen. Rel. and Gravitation 
{\bf 28} (1996) 565. 

\bibitem{banks} T. Banks, M. Berkooz, S.H. Shenker, G. Moore and P.J. 
Steinhardt, Phys. Rev. {\bf D52} (1995) 3548.

\bibitem{linde} A. Linde, Phys. Lett. {\bf B372} (1994) 208.

\bibitem{vilenkin} A. Vilenkin, Phys. Rev. Lett. {\bf 72} (1994) 3137.

\bibitem{ferrara} S. Ferrara, D. L\"ust, A. Shapere and S. Theisen, 
Phys. Lett. {\bf B225} (1989) 363.

\bibitem{sakharov} 
A.D. Sakharov, JETP Lett. {\bf{5}} (1967) 24.

\bibitem{yoshimura} 
M. Yoshimura, 
Phys. Rev. Lett. {\bf 41} (1978) 281; 
Phys. Lett. {\bf B88} (1979) 294.

\bibitem{ig} 
A.Yu. Ignatiev, N.V. Krasnikov, V.A. Kuzmin 
and A.N. Tavkhelidze, Phys. Lett. {\bf B76} (1978) 436.

\bibitem{di} 
S. Dimopoulos and L. Susskind, 
Phys. Rev. D {\bf 18} (1978) 4500. 

\bibitem{we} 
S. Weinberg, Phys. Rev. Lett. {\bf 42} (1979) 850.  

\bibitem{claudson} 
M. Claudson, L.J. Hall and I. Hinchliffe, 
Nucl. Phys. {\bf B241} (1984) 309.

\bibitem{ya} 
K. Yamamoto, Phys. Lett. {\bf B168} (1986) 341.

\bibitem{be} 
O. Bertolami and G.G. Ross, 
Phys. Lett. {\bf B183} (1987) 163.


\bibitem{affleck} 
I. Affleck and M. Dine, 
Nucl. Phys.  {\bf B249} (1985) 361.

\bibitem{kuzmin} 
V.A. Kuzmin, V.A. Rubakov and M.E. Shaposhnikov, 
Phys. Lett. {\bf B155} (1985) 36. 

\bibitem{kuzmin1} 
V.A. Kuzmin, V.A. Rubakov and M.E. Shaposhnikov, 
Phys. Lett.  {\bf B191} (1987) 171.

\bibitem{dolgov} 
A.D. Dolgov and Ya.B. Zeldovich, 
Rev. Mod. Phys. {\bf 53} (1981) 1.

\bibitem{cohen} 
A.G. Cohen and D.B. Kaplan, 
Phys. Lett.  {\bf B199} (1987) 251; 
Nucl. Phys. {\bf B308} (1988) 913.

         
\bibitem{meson}
D. Colladay and V.A. Kosteleck\'y, 
Phys. Lett. {\bf B344} (1995) 259;
Phys. Rev. {\bf D52} (1995) 6224;
V.A. Kosteleck\'y and R. Van Kooten, 
Phys. Rev. {\bf D54} (1996) 5585.

\bibitem{ambjorn}
J. Ambjorn and A. Krasnitz,
Phys. Lett. {\bf B362} (1995) 97.

\bibitem{rs} 
V.A. Rubakov and M.E. Shaposhnikov, 
Uspekhi Fiz.\ Nauk {\bf 166} (1996) 493
(hep-ph/9603208).

\bibitem{yoshimura1}  
M. Yoshimura, 
Phys. Rev. Lett. {\bf 66} (1991) 1559.

\bibitem{bento}  
M.C. Bento, O. Bertolami and P.M. S\'a, 
Mod. Phys. Lett. {\bf A7} (1992) 911.

\bibitem{candelas} P. Candelas, G.T. Horowitz, A. Strominger and E. Witten, 
\NP {\bf B258} (1985) 46.  


\bibitem{hh}J.B. Hartle and S.W. Hawking, \PR {\bf D28} (1983) 2960.

\bibitem{coleman} S. Coleman, \NP {\bf B307} (1988) 867.

\bibitem{veneziano} G. Veneziano, Phys. Lett. {\bf B265} (1991) 287.

\bibitem{tseytlin} A.A. Tseytlin, Mod. Phys. Lett. {\bf 6} (1991) 1721.

\bibitem{bento3} M.C. Bento and O. Bertolami, \CQG {\bf 12} (1995) 1919.  

\bibitem{gasperini} M. Gasperini, J. Maharana and G. Veneziano, \NP 
{\bf B472} (1996) 349.


\bibitem{lukas} A. Lukas and R. Poppe, ``Decoherence in Pre-Big-Bang
Cosmology'' (hep-th/9603167).

\bibitem{bmpv}O. Bertolami, J.M. Mour\~ao, R.F. Picken and I.P. Volobujev, 
\IJMP {\bf A6} (1991) 4149. 

\bibitem{bkm}O. Bertolami, Yu.A. Kubyshin and J.M. Mour\~ao, 
\PR {\bf D45} (1992) 3405. 


\bibitem{bm}O. Bertolami and J.M. Mour\~ao, \CQG {\bf 8} (1991) 1271.


\bibitem{bmon}O. Bertolami and P.V. Moniz, \NP {\bf B439} (1995) 259.


 
\bibitem{jh}J.J. Halliwell, \NP {\bf B266} (1986) 228.


\bibitem{cw}U. Carow-Watamura, T. Inami and S. Watamura, 
\CQG {\bf 4} (1987) 23.



                              
\end{thebibliography}
\end{document}